\documentclass[reprint,nofootinbib,superscriptaddress,amsmath,amssymb,aps,prd]{revtex4-1}
\usepackage{amsmath}
\usepackage{graphicx}
\usepackage{color}
\usepackage{dcolumn}
\usepackage{bm}
\usepackage{slashed}
\usepackage{epstopdf}
\usepackage{multirow}
\usepackage{float}
\usepackage{mathptmx}
\usepackage{enumitem}
\usepackage[colorlinks,linkcolor=red,anchorcolor=green,citecolor=blue,CJKbookmarks=True]{hyperref}
\DeclareSymbolFont{symbols} {OMS}{cmsy}{m}{n}

\newcommand{\bea}{\begin{eqnarray*}}
\newcommand{\eea}{\end{eqnarray*}}
\newcommand{\bean}{\begin{eqnarray}}
\newcommand{\eean}{\end{eqnarray}}

\newcommand{\meq}[1]{(\ref{#1})}
\newcommand{\grad}{\nabla}

\newcommand{\hsp}{\hspace{0.1mm}}

\newcommand{\pp}{\partial}

\begin{document}
\title{Lorentz violation induces isospectrality breaking in Einstein-Bumblebee gravity theory}

\author{Wentao Liu}
\affiliation{Department of Physics, Key Laboratory of Low Dimensional Quantum Structures and Quantum Control of Ministry of Education, and Synergetic Innovation Center for Quantum Effects and Applications, Hunan Normal
University, Changsha, Hunan 410081, P. R. China}

\author{Xiongjun Fang}
\email{fangxj@hunnu.edu.cn} \affiliation{Department of Physics, Key Laboratory of Low Dimensional Quantum Structures and Quantum Control of Ministry of Education, and Synergetic Innovation Center for Quantum Effects and Applications, Hunan Normal
University, Changsha, Hunan 410081, P. R. China}

\author{Jiliang Jing}
\affiliation{Department of Physics, Key Laboratory of Low Dimensional Quantum Structures and Quantum Control of Ministry of Education, and Synergetic Innovation Center for Quantum Effects and Applications, Hunan Normal
University, Changsha, Hunan 410081, P. R. China}

\author{Jieci Wang}
\email{jcwang@hunnu.edu.cn}\affiliation{Department of Physics, Key Laboratory of Low Dimensional Quantum Structures and Quantum Control of Ministry of Education, and Synergetic Innovation Center for Quantum Effects and Applications, Hunan Normal
University, Changsha, Hunan 410081, P. R. China}

\begin{abstract}

In this paper, we investigate the quasinormal modes (QNMs) of a Lorentz-violating spacetime, factoring in a cosmological constant, within the framework of Einstein-Bumblebee gravity. 
Our findings reveal that the interaction of spacetime with an anisotropic bumblebee field imparts distinct contributions to the axial and polar sectors of the vector perturbations. 
This subsequently breaks the isospectrality typically observed in vector modes.
Numerical evidence strongly indicates isospectral breaking in the vector modes of Einstein-Bumblebee black holes: a pronounced breakage in the real part of the frequencies, while the imaginary component seems less affected.
This isospectral breaking indicates the existence of two different waveforms in the Ringdown phase of the black hole, which provides a potential signal of quantum gravity observable in current experiments.

\end{abstract}

\pacs{~}

\maketitle

\section{Introduction}


Lorentz symmetry is a fundamental spacetime symmetry. 
It forms the foundation of quantum field theory and the standard model of modern particle physics, while also playing a crucial role in General Relativity (GR).
Motivated by the pursuit of quantum gravity and evidence from high-energy cosmic rays, researchers have extensively discussed the potential for Lorentz symmetry breaking at the Planck energy scale \cite{Kostelecky1991,Kostelecky1998,Kostelecky19891,Gambini1999,Kostelecky2001}.
By investigating the low-energy contribution from Lorentz symmetry breaking, especially its impact on spacetime compared to GR, and analyzing high-precision observation data, we can explore the possibility of Lorentz symmetry breaking in spacetimes. 
This analysis also allows us to assess whether these implications are consistent with a quantum gravity model, requiring an accurate description of the observed phenomena within that framework.

As a vector-tensor theory and an extension of the Einstein-Maxwell theory, the bumblebee theory incorporates a nonminimally coupled vector field $B_\mu$, known as the bumblebee field. 
This field, when acquiring a nonzero vacuum expectation value (VEV) under an appropriate potential, instigates spontaneous Lorentz symmetry breaking (LSB) and extends the standard general relativity formalism.
The Bumblebee field delineates a privileged direction within spacetime, thereby implying the generation of an anisotropic energy-momentum tensor.
Kostelecky and Samuel were pioneers in introducing the bumblebee gravitational model to study the spontaneous Lorentz violation (LV) effects \cite{Kostelecky1989}.
Later, Casana et al. introduced a Schwarzschild-like bumblebee black hole \cite{Casana2018}. 
Subsequently, various spherically symmetric solutions in bumblebee gravity were found, including wormholes, global monopoles, a cosmological constant, and the Einstein-Gauss-Bonnet term \cite{Ovgun2019,Gullu2020,Maluf2021,Ding2022}.
Ding et al. obtained a Kerr-like solution in Bumblebee gravity \cite{Ding2020} which only approximates the field equations.
Shortly thereafter, Poulis provided a rotation solution that strictly satisfies the field equations \cite{Poulis:2021nqh}.
Based on the aforementioned solutions, the study of Lorentz symmetry breaking effects become an active area in black hole physics \cite{Liu:2022dcn,Mai:2023ggs,Yang:2023wtu,Xu:2023xqh,Zhang:2023wwk,Lin:2023foj,Chen:2023cjd,Chen2020,wang2021}.


On the other hand, the detection of gravitational waves provides a way to verify the consistency of modified theories of gravity and models motivated by quantum gravity theories.
The coalescence of binary black holes, includes three distinct phases: inspiral, merger, and ringdown \cite{Jing2022}.
The ringdown phase is usually described by the perturbation of black hole spacetime, which are called quasinormal modes.
Perturbations of arbitrary spin fields create space-time ripples with complex frequencies that propagate both outward and towards the black hole's horizon.
The frequency's real part indicates time oscillations, while the imaginary part results in exponential decay of the perturbation.
In general relativity, both electromagnetic and gravitational perturbations exhibit axial and polar modes, with their isospectrality demonstrated across various spacetimes: for instance, the Schwarzschild and the Reissner-Nordstr{\"o}m spacetimes \cite{Chandbook}.
Recently, Pani et al. have provided strong evidence for the isospectrality at the first order of rotation of the Kerr metric and Kerr-Newman metric, using numerical results \cite{Pani2012,Pani2012prl,Pani2013prl,Pani2013prd,Pani2013IJMPA}.
A priori, there is no reason to expect that such a remarkable property would also hold true for black holes in some modified theories of gravity.
It is well known that isospectrality can be easily broken: for example, axial and polar modes are not isospectral (even for nonrotating black holes) in higher-dimensional scenarios \cite{Berti2009}, or when the underlying theory deviates from general relativity, as in the cases of loop quantum gravity \cite{isoLoop} and Chern-Simons gravity \cite{isoChSi}.



This paper is dedicated to exploring the effects of Lorentz symmetry breaking in Einstein-Bumblebee black hole spacetimes on the spectral properties of vector modes.
We look forward to establishing a connection between Lorentz violation and the possibility of isospectral breaking in gravitational wave observations.
The direct detection of these frequencies is expected to be possible with sufficient precision in current gravitational wave detectors, such as LIGO and Virgo \cite{Isi:2019aib}.
The paper is organized as follows. 
In Section \ref{Sec.2}, we briefly review the spherically symmetric black holes in Einstein-Bumblebee gravity theory and derive the Schr\"{o}dinger-like equations for the (axial and polar) vector perturbations. 
Section \ref{Sec.3} focuses on the study of QNM frequencies and the ringdown waveforms of the vector modes. 
Using the matrix method and the continued fraction method, we numerically calculate the QNM frequencies, and the accuracy of the results is confirmed by fitting the waveforms.
Additionally, we present a comparative discussion regarding the QNM frequency splitting phenomenon.
Sec. \ref{Sec.4} is dedicated to summarizing our results and discussing the foreseeability of isospectral violation of gravitational perturbations in this theory.
Throughout our paper, we adopt the conventions $ c=G=1 $ and use the factor $ \kappa=8\pi $ in our numerical calculation.

\section{Formalism}\label{Sec.2}

\subsection{Black hole solutions}

The action for the bumblebee field $B_\mu$ coupled to gravity can be described as \cite{Casana2018}
\begin{equation}
\begin{aligned}\label{Action}
\mathcal{S}_B=&\int d^4x \sqrt{-g}\left[\frac{1}{2\kappa}\left(R-2\Lambda\right)+\frac{\zeta}{2\kappa} B^\mu B^\nu R_{\mu\nu} \right.\\
&\left. -\frac{1}{4}B^{\mu\nu}B_{\mu\nu}-V\left(B^\mu B_\mu\pm b^2\right)+\mathcal{L}_A\right],
\end{aligned}
\end{equation}

where $ \kappa=8\pi G_N $ is the gravitational coupling constant that can be set to $ G_N =1$ without loss of generality.
$ \Lambda $ is the cosmological constant and $ \zeta $ is the real coupling constant which controls the non-minimal gravity interaction to the bumblebee field $B_\mu$. 
The strength of this field is defined by $ B_{\mu\nu}=\pp_\mu B_\nu-\pp_\nu B_\mu$.
It is worth noting that the potential $V$, chosen to ensure a non-zero VEV for the bumblebee field $\langle B_\mu\rangle\equiv b_\mu$, triggers spontaneous Lorentz symmetry breaking, reaching its minimum at $B_\mu B^\mu \pm b^2$, where $b$ is a positive real constant,  and the $ \pm $ sign implies that $ B_\mu $ is timelike or spacelike, respectively.
$ \mathcal{L}_A $ describes a vacuum excitation vector perturbation field, which maintains the same form as the Bumblebee field, and is given by
\begin{align}
\mathcal{L}_A=\frac{\xi}{2\kappa} A^\mu A^\nu R_{\mu\nu}-V(A^\mu A_\mu) -\frac{1}{4}A^{\mu\nu}A_{\mu\nu}.
\end{align}
Its contribution to the gravitational field is negligible.

To investigate the effect of the Lorentz violation when a cosmological constant exists, one could consider relaxing the vacuum conditions, $ V=0 $ and $ V'=0 $, originally assumed by Casana et al \cite{Casana2018}.
A prime exemplar of a potential that fulfills these conditions is a smooth quadratic function, denoted as
\begin{align}
V=V(X)=\frac{\lambda}{2}X^2,
\end{align}
where $ \lambda $ serves as a constant, and $ X $ signifies a generic potential argument. 
Consequently, the VEV, $ b_\mu $, emerges as a solution of $ V=V'=0$.
Another simple choice of potential consists of a linear function assumed by Maluf \cite{Maluf2021}
\begin{align}
V=V(\lambda,X)=\frac{\lambda}{2}X.
\end{align}
In this context, $ \lambda $ is interpreted as a Lagrange-multiplier field \cite{Bluhm:2007bd}. Its equation of motion ensures the vacuum condition $ X=0 $, leading to $ V=0 $ for any field $ \lambda $ on-shell.
Notably, this form of potential manifests itself as a cosmological constant. 
It is this latter assumption that interests us, namely
\begin{align}\label{VVV}
&V(B^\mu B_\mu -b^2)=\frac{\lambda}{2}(B^\mu B_\mu -b^2)=0,\\ \label{Vp}
&V'(B^\mu B_\mu -b^2)=\frac{\lambda}{2},
\end{align}
where $ V'(X)=dV(X)/dX $.

Toking of the variational $g_{\mu\nu}$ and $B_\mu$ yields the gravitational equation and bumblebee field equation:
\begin{align}
& R_{\mu\nu}-\frac{1}{2}g_{\mu\nu}\left(R-2\Lambda\right)=\kappa T^{B}_{\mu\nu} \label{EinsteinEQ}, \\ \label{VEq}
& \nabla^\mu B_{\mu\nu}=2V'B_\nu-\frac{\zeta}{\kappa}B^\mu R_{\mu\nu}.
\end{align}
 $T^{B}_{\mu\nu}$ is the bumblebee energy momentum tensor, which have the following form:
\begin{equation}
\begin{aligned}\label{TBab}
T^{B}_{\mu\nu}=
& B_{\mu\alpha}B^\alpha\hsp_\nu-\frac{1}{4}g_{\mu\nu}B^{\alpha\beta}B_{\alpha\beta}-g_{\mu\nu}V+2B_\mu B_\nu V' \\
&+\frac{\zeta}{\kappa}
\left[\frac{1}{2}g_{\mu\nu}B^{\alpha}B^{\beta}R_{\alpha\beta}-B_\mu B^\alpha R_{\alpha \nu}-B_\nu B^\alpha R_{\alpha \mu}\right.\\
& \left.+\frac{1}{2}\nabla_\alpha\nabla_\mu\left(B^\alpha B_\nu\right)
+\frac{1}{2}\nabla_\alpha\nabla_\nu\left(B^\alpha B_\mu\right)\right.\\
&\left.-\frac{1}{2}\nabla^2\left(B_\mu B_\nu\right)-\frac{1}{2}g_{\mu\nu}\nabla_\alpha\nabla_\beta\left(B^\alpha B^\beta\right)  \right].
\end{aligned}
\end{equation}

Note that we ignore the effect of the perturbation field $ A^\mu $ on the background spacetime.

An exact spherically symmetric black hole solution with a cosmological constant has been constructed by Maluf et al. \cite{Maluf2021}, and it is referred to as the Schwarzschild (anti-) de Sitter-like black holes.
The radial bumblebee field $ B_\mu $ can be written as
\begin{align}
B_\mu=b_\mu=(0,b\sqrt{(1+\ell)/F(r)},0,0),
\end{align}
and the metric is given 
\begin{align}\label{ds2MOG}
g_{\mu\nu}=&\text{diag}\left(-F(r),(1+\ell)/F(r),r^2,r^2 \sin^2\theta \right),\\
&F(r)=1-\frac{2M}{r}-(1+\ell)\frac{\Lambda_e}{3}r^2,
\end{align}
where, we regard $ \Lambda_e=\kappa\lambda/\zeta $ as an effective cosmological constant, and the symbol $ \ell=\zeta b^2 $ represents the Lorentz violation parameter.
Note that given the potential described by \meq{VVV} and \meq{Vp}, a solution of this type will only be possible if 
\begin{align}\label{Lambdae}
\Lambda=(1+\ell)\Lambda_e.
\end{align}
The constraint \meq{Lambdae} is an indispensable prerequisite for generating a metric with a cosmological constant from the potential \meq{VVV}-\meq{Vp} and the modified Einstein equations \meq{EinsteinEQ}.
This is the constraint on the field $ \lambda $ derived from the modified Einstein field equations, as previously discussed in \cite{Maluf2021}.
The function \( F(r) \) can be expressed in terms of the black hole horizons as follows:
\begin{align}
F(r)=(1+\ell)\frac{\Lambda_e}{3}\left(1-\frac{r_h}{r}\right)\left(r_c-r\right)\left(r+r_h+r_c \right),
\end{align}
where
\begin{align}
\Lambda_e=\frac{6M}{r_h r_c(r_h+r_c)(1+\ell)},
\end{align}
and \( M \), representing the black hole mass, is given by
\begin{align}
M=\frac{r_h r_c(r_h+r_c)}{2(r_h^2 +r_h+r_c+r_c^2)}.
\end{align}
Here, \( r_h \) and \( r_c \) respectively represent the event horizon and cosmological horizon.
From a gravitational perspective, such solution are supported by an anisotropic energy-momentum tensor, considered as the manifestation of the bumblebee field within spacetime geometry.

\subsection{Linear perturbation}
Black hole linear perturbations play a crucial role in many astrophysical processes, viewed as minor deviations from the background spacetime. 
For example, QNMs aptly describe late stages of binary black hole mergers or gravitational collapses. 
To calculate QNMs, it is usually necessary to obtain a decoupled second-order differential equation in the frequency domain.
Given that the background is spherically symmetric, projecting the perturbations on the basis of scalar, vector, and tensor spherical harmonics allows for the automatic decoupling of the angular and radial parts, as well as the automatic separation of the axial and polar sections of the perturbations.
Let's introduce the spherical harmonic basis notations \cite{Thorne1980,Thompson2016,guage2022}. 
We first define two orthogonal, unnormalized co-vectors $v$ and $n$ as
\begin{equation}
v_\mu=(-1,0,0,0), \quad\quad n_\mu=(0,1,0,0),
\end{equation}
the projection operator
\begin{equation}
\Omega_{\mu\nu}=r^2 \text{diag}(0,0,1,\sin^2\theta),
\end{equation}
and the spatial Levi-Civita tensor, $\epsilon_{\mu\nu\alpha}\equiv v^\beta\epsilon_{\beta \mu \nu \alpha}$, where $\epsilon_{tr\theta\varphi}=r^2\sin\theta$. 
Using the scalar spherical harmonics $Y^{lm}=Y^{lm}(\theta,\varphi)$, the pure-spin vector harmonics are given by
\begin{equation}
\begin{aligned}\label{vectorbasis}
Y_\mu^{E,lm}&= r\grad_\mu Y^{lm}, \\
Y_\mu^{B,lm}&= r\epsilon_{\mu\nu}\hsp^\alpha n^\nu\grad_\alpha Y^{lm}, \\
Y_\mu^{R,lm}&= n_\mu Y^{lm},
\end{aligned}
\end{equation}
and the pure-spin tensor harmonics can be find in \cite{guage2022,Liu:2023uft}.

These spherical harmonic basis will be used to decompose vector or tensor perturbations.
With the background space-time, we can consider the perturbation equation for the vector field $ A_\mu $, given by
\begin{align}
& \nabla^\mu A_{\mu\nu}+\frac{\xi}{\kappa}A^\mu R_{\mu\nu}-\lambda A_\nu=0.
\end{align}
Generally speaking, perturbed vector potential can be written as
\begin{equation}
\begin{aligned}
A_\mu=& \mathrm{P}_{lm}v_\mu Y^{lm}+\mathrm{R}_{lm}Y^{R,lm}_\mu\\
&+\mathrm{S}_{lm}Y^{E,lm}_\mu+\mathrm{Q}_{lm}Y^{B,lm}_\mu.
\end{aligned}
\end{equation}
It is worth noting that the dipole mode ($ l=1 $) is expected to be the most astrophysically relevant; in this scenario, the dynamic behavior of the $ A_\mu $ field and the bumblebee perturbation field $ \delta B_\mu $  is consistent.
Physically, the decoupling can be attributed to the nonradiative nature of dipolar gravitational fields, with the quadrupole ($ l=2 $) being the first radiative multipole. 
For example, in GR, electromagnetic perturbations with dipole modes are always decoupled from gravitational perturbations \cite{Berti:2005eb}.
Mathematically, this decoupling is applicable to any modified gravitational theory that employs spherical harmonics for its multi-mode expansion \cite{Liu:2023uft}.

Then, by separating the angular component from the perturbation equations, we derive a pure radial equation in the following form:
\begin{equation}
\begin{aligned}\label{EQE1}
&\left[2\kappa(1+\ell)\left(l^2+l-r^2\lambda\right)-r\xi\left(2F'+rF''\right)\right]\mathrm{P}_{l}\\
&+2i\kappa\omega rF\left(2\mathrm{R}_l+r\mathrm{R}'_l\right)-2\kappa r F\left(2\mathrm{P}'_l+r\mathrm{P}''_l\right)=0,
\end{aligned}
\end{equation}
\begin{equation}
\begin{aligned}\label{EQE2}
&\left[2\kappa(1+\ell)\left(l^2+l-r^2\lambda\right)F-r\xi F\left(2F'+rF''\right)\right]\mathrm{R}_l\\
&-2\kappa (1+\ell)\omega^2r^2\mathrm{R}_l-2i\kappa (1+\ell)\omega r^2\mathrm{P}'_l=0,
\end{aligned}
\end{equation}
\begin{equation}
\begin{aligned}\label{EQE3}
-i\omega (1+\ell)\mathrm{P}_l+F'F\mathrm{R}_l+F^2\mathrm{R}'_l=0,
\end{aligned}
\end{equation}
\begin{equation}\label{EQQ}
\begin{aligned}
&\left[\kappa (1+\ell)\left(l^2+l-r^2\lambda+\xi/\kappa-\omega^2r^2/F\right)-\xi F\right]\mathrm{Q}_l\\
&-r(\kappa+\xi)F'\mathrm{Q}_l-r\kappa\left(2F+rF'\right)\mathrm{Q}'_l-r^2\kappa F \mathrm{Q}''_l=0.
\end{aligned}
\end{equation}
The perturbations can be naturally divided into two sections: the polar sector, encompassing equations \eqref{EQE1} through \eqref{EQE3}, and the axial sector, described by equation \eqref{EQQ}.
Note that we consider a time dependence of each perturbation functions, e.g., $ \mathrm{P}_{l}(t,r)=e^{-i \omega t} \mathrm{P}_{l}(r) $, and fix the gauge such that $ \mathrm{S}_l=0 $.

To obtain the Schr\"{o}dinger-like equations, we eliminate the perturbation function \( \mathrm{P}_l \) from equation \eqref{EQE3} and perform a coordinate transformation
\begin{align}
\psi_l^{+}(r)=F\mathrm{R}_l(r),~~~ \text{and} ~~~\psi^{-}_l(r)=\frac{1}{r} \mathrm{Q}_l(r).
\end{align}
The polar ``$ + $'' and axial ``$ - $'' parity oscillations can be obtained using equations \eqref{EQE2} and \eqref{EQQ}, respectively,
\begin{align}\label{masterEq}
\frac{d^2}{dr_*^2}\psi^{\pm}_l+\left[(1+\ell)\omega^2-V^{\pm}_l\right]\psi^{\pm}_l=0,
\end{align}
where \( r_*=\int \frac{1}{F(r)}dr \) is the tortoise coordinate and the effective potentials are given by
\begin{equation}
V^{\pm}_l=\frac{F}{r^2}
\left(
\begin{array}{c}
 (1+\ell)\left(l^2 +l-r^2\lambda+r^2 \Lambda_e\xi/\kappa  \right) \\
 (1+\ell)\left(l^2 +l-r^2\lambda+r^2 \Lambda_e\xi/\kappa  \right)+\ell\xi/\kappa
\end{array}
\right).
\end{equation}
Since the Bumblebee fields provides a privileged direction in spacetime, imparting distinct contributions to the polar and axial sectors of the vector perturbations. 
This leads to the existence of different effective potentials for the two parities.
Here the tortoise coordinate $r_*$ can be rewritten as
\begin{align}
r_*=&\eta_h\ln\left(r-r_h\right)+\eta_c \ln \left(r_c-r\right)+\eta_i\ln\left(r+r_h+r_c\right),
\end{align}
where $ \eta_i=-\eta_h-\eta_c $ and,
\begin{equation}
\begin{aligned}
\eta_h&=\frac{3r_h}{(1+\ell)\Lambda_e}(r_c-r_h)^{-1}(r_h+2r_c)^{-1},\\
\eta_c&=\frac{3r_c}{(1+\ell)\Lambda_e}\left(r_h^2+r_h r_c-2r_c^2\right)^{-1}.
\end{aligned}
\end{equation}

The equations \meq{masterEq} satisfy two boundary conditions, owing to the existence of the two horizons. 
For simplicity, we set \( \zeta=\xi \); and then the generic wave function \(\psi_l^{\pm}\) exhibits the following asymptotic behaviors:
\begin{equation}\label{solveeq}
\psi_l^{\pm}\sim
\begin{cases}
e^{-i\sqrt{1+\ell}\omega r_*} &\text{for}~~r\rightarrow~r_h,\\
e^{i\sqrt{1+\ell}\omega  r_*} &\text{for}~~r\rightarrow~r_c ,
\end{cases}
\end{equation}
which indicates an pure-ingoing wave at the event horizon and an outgoing wave at the cosmological horizon.
Here, the eigenvalue $ \omega $ is designated as the QNM frequencies, typically characterized as a complex number. 
This is attributed to the dissipative nature of the boundary condition equation \meq{solveeq}, leading to a non-self-adjoint differential operator for this system.
For convenience, we define $ \tilde{\omega}=\sqrt{1+\ell}\omega $ as the effective frequency.

\section{Numerical results}\label{Sec.3}

\subsection{Numerical methods}
In this subsection, we calculate QNMs with the matrix method (MM), which was proposed by Lin et al. \cite{Lin2017,Lin20171,Lin2019,Lei2021}.
And we also use the continued fraction method (CFM) which is groundbreaking work by Leaver \cite{Leaver1985} in order to verify the QNMs frequencies obtained by MM. 

First we briefly review the MM, which is a non-grid-based interpolation approach which can be used to solve the eigenvalue problem.
The perturbation equations and its boundary conditions can be transformed into a homogeneous matrix equation by discretizing the linear partial differential equation \cite{Lin2017}.
Grounded in the information about $ N $ scattered data points, we can construct Taylor series expansions for the unknown eigenfunction up to the $ N $th order at each discretized point.
Subsequently, the derived homogeneous system of linear algebraic equations is resolved to ascertain the eigenvalue.
The main feature that differs from other methods is that the series unwinding of the wave function is carried out near a series of points that are not necessarily evenly distributed in the domain of the wave function.
Through the modulation of interpolation points, one can achieve a desirable level of precision, maintaining a reasonable equilibrium with computational efficiency. 

To convert the radial interval into $ [0, 1] $, we introduce a coordinate transformation
\begin{equation}
x=\frac{r-r_h}{r_c-r_h},
\end{equation}
which is ensured by the outer communication domain $r_h\leq r\leq r_c$.
Together with the boundary conditions, which ensure that $ \chi(0)=\chi(1)=0 $, we consider that $ \psi_l^{\pm} $ satisfy the following relation:
\begin{align}\label{psi}
\psi_l^{\pm}(x)=x(1-x)(1-x)^{i \tilde{\omega}\eta_c}x^{- i\tilde{\omega} \eta_h} \chi(x).
\end{align}
Subsequently, it is feasible to reformulate the perturbation equations for all effective potentials into the ensuing structure:
\begin{equation}\label{qicieq}
\tau_0^{\pm}(x)\chi''(x)+\lambda_0^{\pm}(x)\chi'(x)+s_0^{\pm}(x)\chi(x)=0.
\end{equation}
Numerically, the coefficients $ \tau_0^{\pm}(x) $, $ \lambda_0^{\pm}(x) $, and $ s_0^{\pm}(x) $ are determined by the effective potentials, the coupling constant $ \xi $, as well as the black hole horizons $ r_h $ and $ r_c $, which are determined by the parameters $ M $, $ \Lambda_e $, and $ \ell $.

It becomes necessary to discretize Eq. \meq{qicieq} and incorporate uniformly distributed grid points within the interval $[0,1]$.
The corresponding differential matrices can be constructed by expanding the function $ \chi(x) $ around each grid point using the Taylor series.
Thus, the differential Eq. \meq{qicieq} is therefore rewritten as an algebraic equations take the form as
\begin{align}\label{algeq}
\left(M^{\pm}_0+\sqrt{1+\ell}\omega M_1^{\pm}\right)\chi(x)=0,	
\end{align}
where $ M_0^{\pm} $ and $ M_1^{\pm} $ are matrices consisting of the coefficient functions $ \tau_0^{\pm}(x) $, $ \lambda_0^{\pm}(x) $, and $ s_0^{\pm}(x) $ and the corresponding differential matrices.
By solving algebraic equations $|M^{\pm}_0+\sqrt{1+\ell}\omega M_1^{\pm} |=0$, we can easily obtain the QNM frequencies.

On the other hand, to verify, we use the CFM to calculate QNMs by numerically solving a three-term recurrence relation.
It is convenient to introduce a new independent variable, defined by $ y=1/r $, $ y_h=1/r_h $ and $ y_c=1/r_c $.
With this new variable $ y $, the asymptotic form of the perturbations as $ r_*\rightarrow\infty $ can be rewritten as
\begin{align}\label{psiinf}
e^{i\tilde{\omega}r_*}=\left(y-y_h\right)^{i\tilde{\omega}\eta_h}\left(y-y_c\right)^{i\tilde{\omega}\eta_c}
\left(y+\frac{y_h y_c}{y_h+y_c}\right)^{i\tilde{\omega}\eta_i}.
\end{align}
The perturbation function $ \psi_l^{\pm} $ that expands at the black hole horizon can be written in the following form:
\begin{align}\label{Leaverpsi}
\psi_l^{\pm}=e^{i\tilde{\omega}r_*}\left(y-y_h\right)^{-2i\tilde{\omega}\eta_h} \sum_{n=0}^{\infty}a_n\left(\frac{y-y_h}{y_c-y_h}\right)^n.
\end{align}
For the equation \meq{masterEq}, the expansion coefficients are defined by a three-term recursion relations
\begin{equation}
\begin{aligned}\label{ditui}
\alpha_0^{\pm}a_1&+\beta_0^{\pm}a_0=0,\\
\alpha_n^{\pm}a_{n+1}+\beta_n^{\pm}a_n&+\gamma_n^{\pm}a_{n-1}=0,~~~ n>0.
\end{aligned}
\end{equation}
The recurrence coefficients $ \alpha_n^{\pm} $, $ \beta_n^{\pm} $, and $ \gamma_n^{\pm} $ are simple functions consisting of $ n $ and black hole parameters, which take the explicit forms as follows
exptitat forms as 
\begin{widetext}
\begin{align}
&\begin{aligned}
\alpha_n^{\pm}=&2i\tilde{\omega} r_cr_h(1+n)(r_c+2r_h)^2\left(r_c^3+2r_h^3\right)-r_c(1+n)^2(r_c+2r_h)^3(r_c-r_h)^2,
\end{aligned}\\
&\begin{aligned}
\beta_n^{+}=&\left[n\left(r_c^2+r_cr_h-2r_h^2\right)-2i\tilde{\omega} r_h\left(r_c^2+r_cr_h+r_h^2\right)\right]\left[(n+1)\left(r_c^2+r_cr_h-2r_h^2\right)-2i\tilde{\omega} r_h\left(r_c^2+r_cr_h+r_h^2\right)\right]\\
&\times \left(2r_c^2+2r_cr_h-r_h^2\right)+\left(r_c^2+r_cr_h-2r_h^2\right)^2\left(r_c^2+r_cr_h+r_h^2\right)\left(l^2+l\right)(1+\ell),
\end{aligned}\\
&\begin{aligned}
\beta_n^{-}=\beta_n^{(+)}+(r_c+r_h)^2(r_c+2r_h)^2\left(r_c^2+r_cr_h+r_h^2\right)\ell\frac{\xi}{\kappa},
\end{aligned}\\
&\begin{aligned}
\gamma_n^{\pm}=\left(r_c^2-r_h^2\right)\left[\left(1-n^2\right)(r_c-r_h)^2(r_c+2r_h)^2+4\tilde{\omega}^2 r_h^2\left(r_c^2+r_cr_h+r_h^2\right)^2
+4ni\tilde{\omega} r_h(r_c-r_h)(r_c+2r_h)\left(r_c^2+r_cr_h+r_h^2\right)\right].
\end{aligned}
\end{align}
\end{widetext}
Note that when $ \ell=0 $, the recurrence relation does not return to the Schwarzschild-de Sitter case presented in Ref. \cite{Yoshida:2003zz}, because we chose different time-dependent factors.
Upon comparing the expanded eigenfunction from equation \meq{Leaverpsi} with equation \meq{psiinf}, it becomes clear that the eigenfunction adheres to the QNM boundary condition as defined by \meq{solveeq}, provided that the power series in \meq{Leaverpsi} converges within the range $ y_c \leq y\leq y_h $.
This convergence condition implies that we only need to provide the initial value of $ a_0 $ and recursively calculate at significantly large values of $ n $, where the following condition exists,
\begin{align}\label{ana1}
a_{n}-a_{n-1}=0.
\end{align}
The QNM frequencies can be easily obtained by solving the algebraic equation \meq{ana1}, and we can set the initial value $ a_0=1 $ without loss of generality.

\subsection{Quasinormal modes}
In this subsection, the parameters are set as $ M=1 $ and $ \Lambda_e=0.01 $.
We aim to investigate the effect of the Lorentz violation parameter $ \ell $ and the coupling constant $ \xi $ on the isospectrality of the QNM spectrum.
Within MM, we set $ N = 20 $, to ensure that the relative error becomes smaller than $ 10^{-5} $.
In CFM, considering computational efficiency, the determination is made to utilize the result of $ n=12 $, following a comparative assessment of the cases for $ n=12 $ and $ n=20 $, with the observation that the relative error gravitates towards $10^{-5}$.

The vector modes exist for $ l\geq 1 $.
Our analysis specifically targets modes with azimuthal indices $ l=1 $ and $ l=2 $ due to their astrophysically longer lifetimes and dominance in practical detections.
\begin{figure}[h]
\centering
\includegraphics[width=0.49\linewidth]{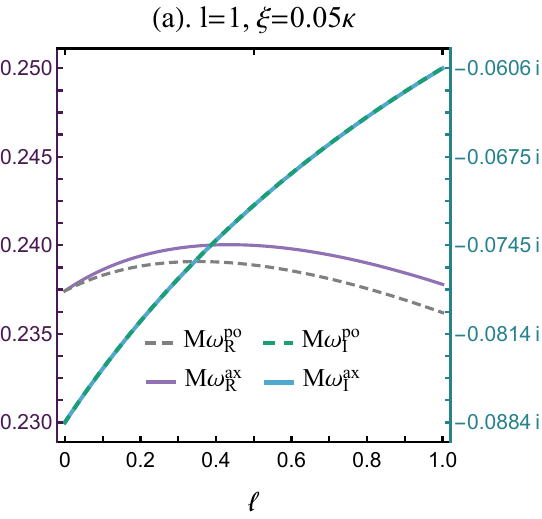}
\includegraphics[width=0.47\linewidth]{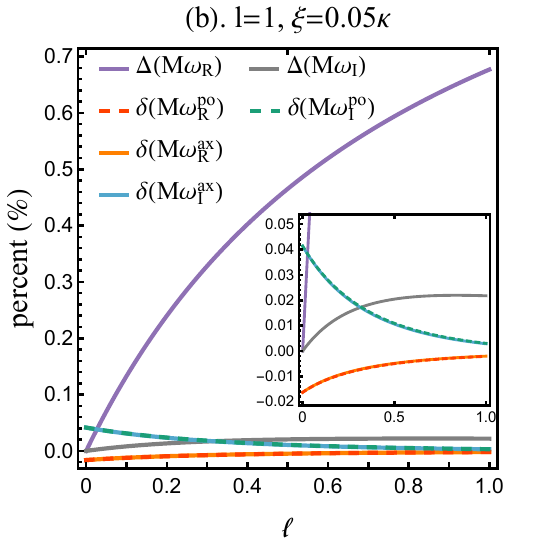}
\includegraphics[width=0.49\linewidth]{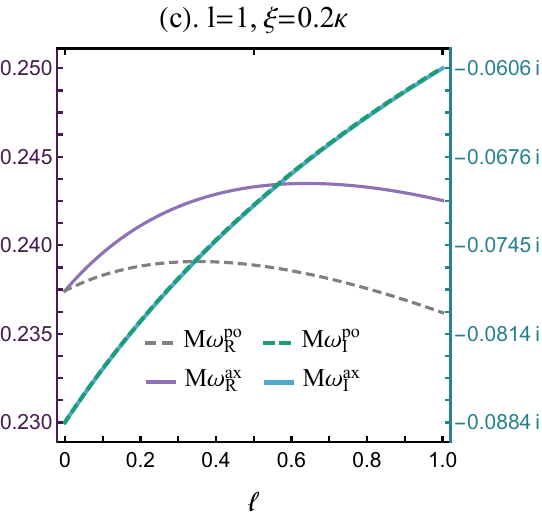}
\includegraphics[width=0.47\linewidth]{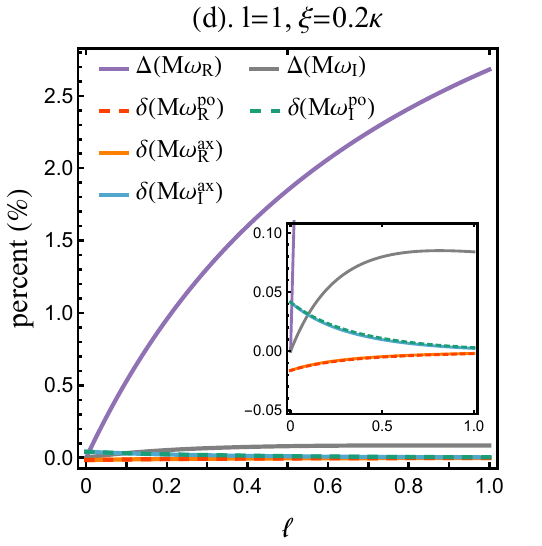}
\caption{The vector frequencies in the $l=1, n=0$ mode for varying values of $ \ell $ are shown in the graph, which correspond to the results when the black hole mass $M=1$.}
\label{fig1}
\end{figure}
In Figs. \ref{fig1}-(a) and \ref{fig1}-(c), we show the fundamental mode of the azimuthal index \( l = 1 \).
For both axial and polar modes, as the Lorentz violation parameter $ \ell $ increases, the real part of the QNM initially increases and then decreases, while the imaginary part increases monotonically.
With a non-zero Lorentz violation parameter, the real part of the QNM spectrum shows a noticeable deviation between axial and polar modes.
Shown in Figs. \ref{fig1}-(b) and \ref{fig1}-(d), the deviation in the imaginary part is merely close to the numerical error between the MM and CFM methods. 
Here, the relative deviation between the two sections is indicated with the symbol $ \Delta $, whereas the relative error between the two numerical methods is indicated with the symbol $ \delta $.
For example,
\begin{equation}
\begin{aligned}
\Delta\left(M\omega_R\right)=& 100\times \frac{M\omega^\text{ax}_\text{R}-M\omega^\text{po}_\text{R}}{M\omega^\text{po}_\text{R}},\\
\delta\left(M\omega^\text{ax}_\text{R}\right)=& 100\times \frac{M\omega^\text{ax}_\text{R,MM}-M\omega^\text{ax}_\text{R,CFM}}{M\omega^\text{ax}_\text{R,CFM}}.
\end{aligned}
\end{equation}
\begin{figure}[h]
\centering
\includegraphics[width=0.49\linewidth]{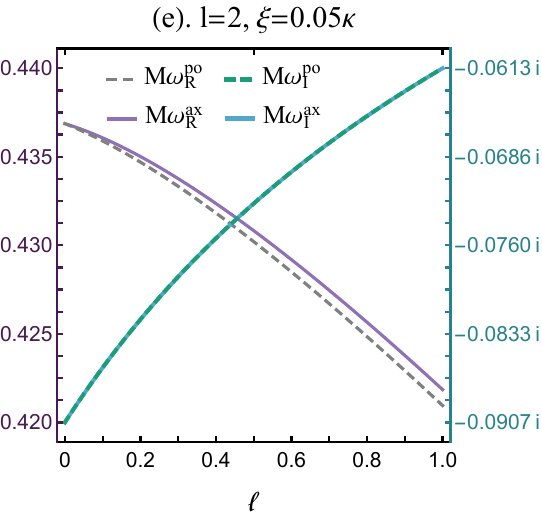}
\includegraphics[width=0.47\linewidth]{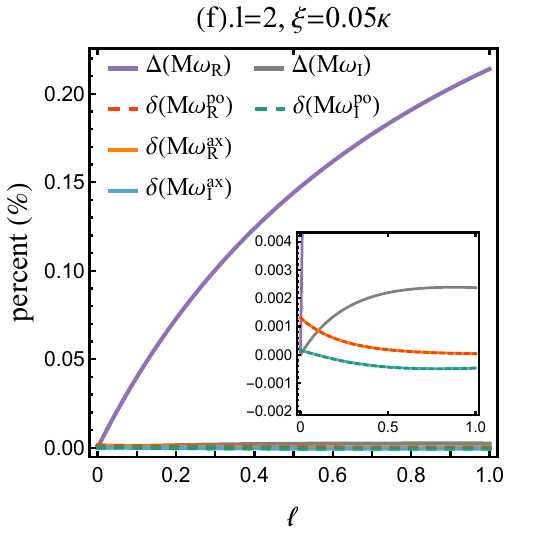}
\includegraphics[width=0.49\linewidth]{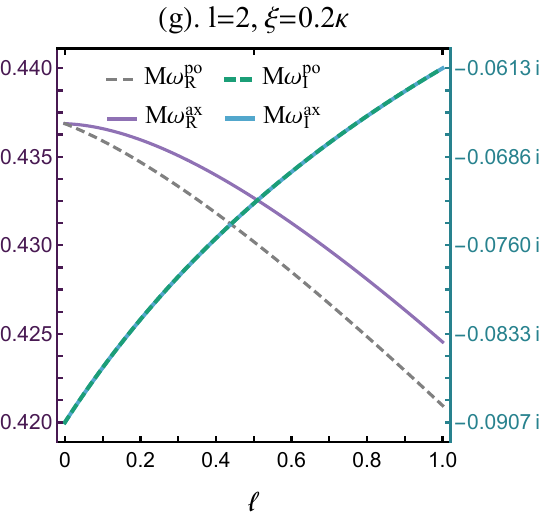}
\includegraphics[width=0.47\linewidth]{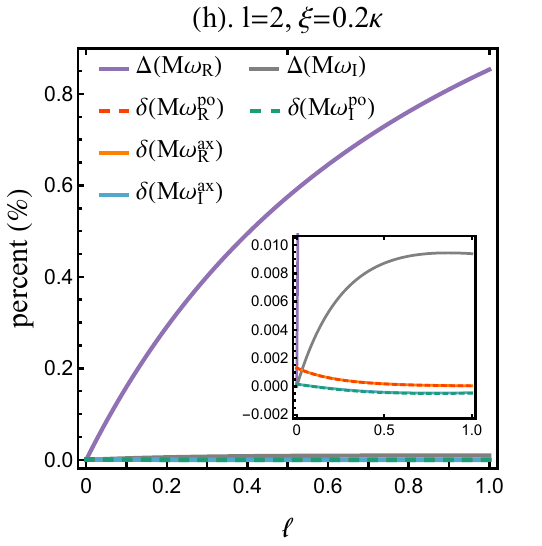}
\caption{The vector frequencies in the $l=2, n=0$ mode for varying values of $ \ell $ are shown in the graph, which correspond to the results when the black hole mass $M=1$.}
\label{fig2}
\end{figure}
The fundamental mode for azimuthal indices $ l = 2 $ is shown in Figs. \ref{fig2}-(e) to \ref{fig2}-(h). 
With an increase in the Lorentz violation parameter $ \ell $, the real part of the QNM rises, while the imaginary part grows monotonically, and the isospectrality breaking still persists.
Additionally, it is noteworthy that the isospectral break is influenced by both the coupling parameter $ \xi $ and the azimuthal number $ l $; it increases with the former and decreases with the latter.

\subsection{Dynamical evolution}
In this subsection, we aim to explore the dynamical evolution of the vector field $ A^\mu $ for a given initial wave packet governed.
The emission of gravitational waves is often accompanied by an electromagnetic counterpart. 
We are particularly interested in the behavior of these vector fields during the ringdown phase. 
And we can validate the conclusions of the previous subsection by comparing the waveforms of the axial and polar modes.
For this, one could perform numerical simulations to solve the perturbation equations for the black holes.

In a finite time domain, we can consider the time-dependent Schr\"{o}dinger-like equation. 
Expressing Eq. \meq{masterEq} as
\begin{align}\label{mastereq}
\frac{\pp^2}{\pp r_*^2}\psi_l^{\pm}-\frac{\pp^2}{\pp t_*^2}\psi_l^{\pm}-V_l^{\pm}\psi_l^{\pm}=0,
\end{align}
where $ t_* = t/\sqrt{1+\ell} $, and using the light-cone coordinates $ u = t_*-r_* $ and $ v = t_* + r_* $, the above equation can be rewritten as
\begin{align}\label{masterequv0}
4\frac{\pp^2 \psi_l^{\pm}(u,v)}{\pp u\pp v}-V_l^{\pm}(u,v)\psi^{\pm}_l(u,v)=0.
\end{align}
Particularly, equation \meq{mastereq} has transformed into a wave equation in \(u\) and \(v\). 
It can be solved using efficient numerical techniques, such as the finite difference method (FDM).
Given the initial distribution $ \psi^{\pm}_l(u,v) $ by imposing the following initial conditions
\begin{align}\label{masterequv}
\psi^{\pm}_l(u,0)=0,~~\psi^{\pm}_l(0,v)=\text{Exp}\left(-\frac{(v-v_c)^2}{2\sigma^2}\right),
\end{align}
where $ \psi^{\pm}_l(0,v) $ denotes a Gaussian wave packet, localized around $v_c$ with a $\sigma$ width.
The observer is positioned at $r_0=10 r_h$ in the Boyer-Lindquist coordinates, situated in the outer communication domain and satisfying the condition $r_h < r_0 < r_c$. 
We then obtain the ringdown waveforms by numerically solving the partial differential equation \meq{masterequv}.
\begin{figure}[h]
\centering
\includegraphics[width=0.47\linewidth]{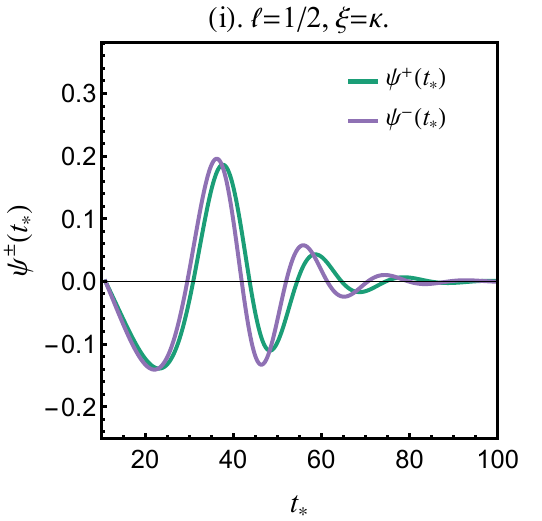}
\includegraphics[width=0.48\linewidth]{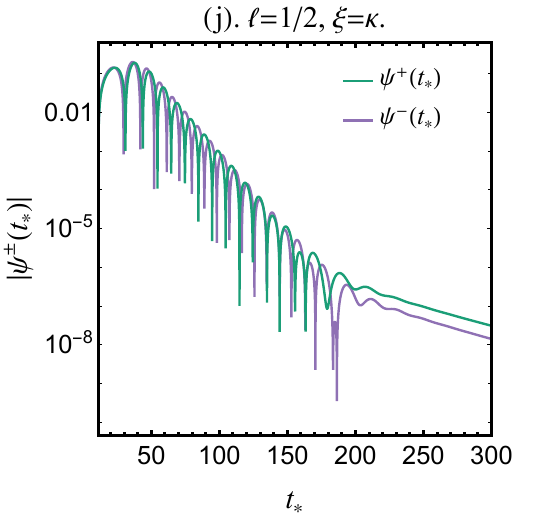}
\includegraphics[width=0.47\linewidth]{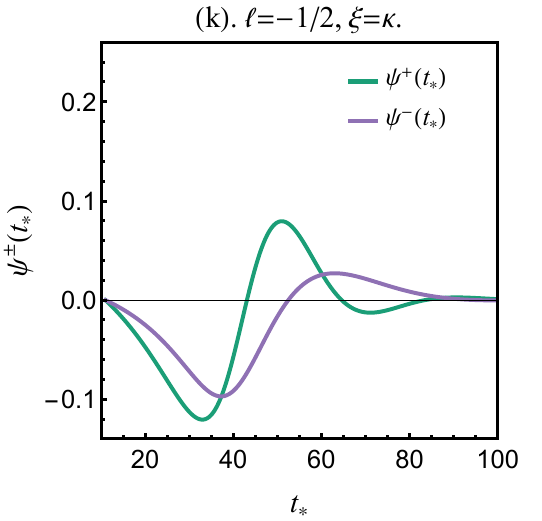}
\includegraphics[width=0.49\linewidth]{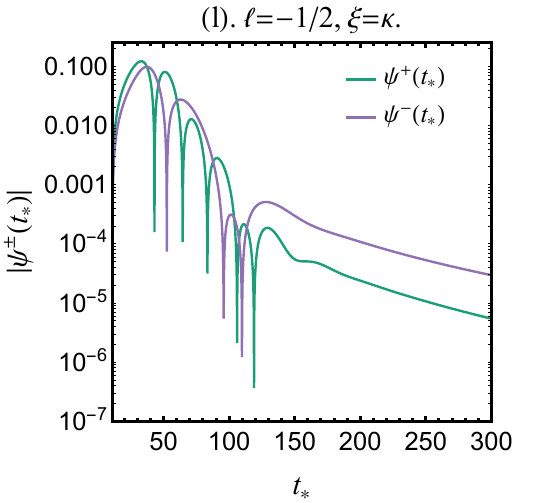}
\caption{The time evolution of the wave function $ \psi^{\pm}(t_*) $ corresponds to the vector perturbations in the $ l=1, n=0 $ mode.}
\label{fig3}
\end{figure}
\begin{figure}[h]
\centering
\includegraphics[width=0.47\linewidth]{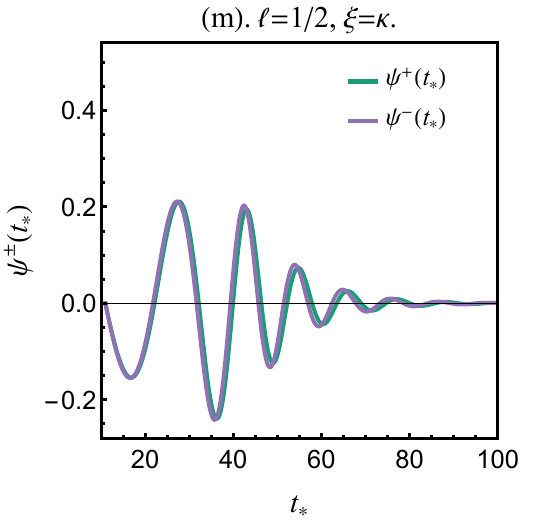}
\includegraphics[width=0.49\linewidth]{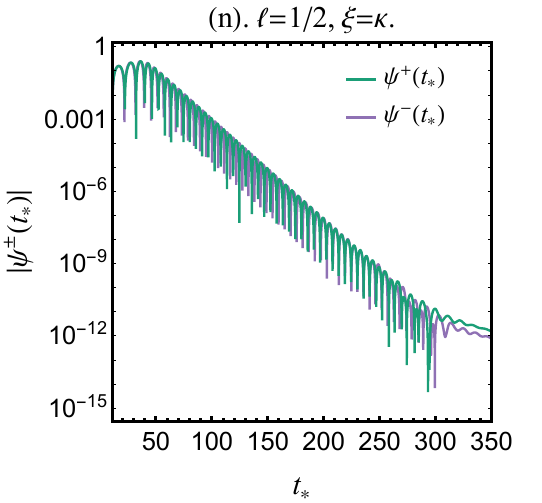}
\includegraphics[width=0.47\linewidth]{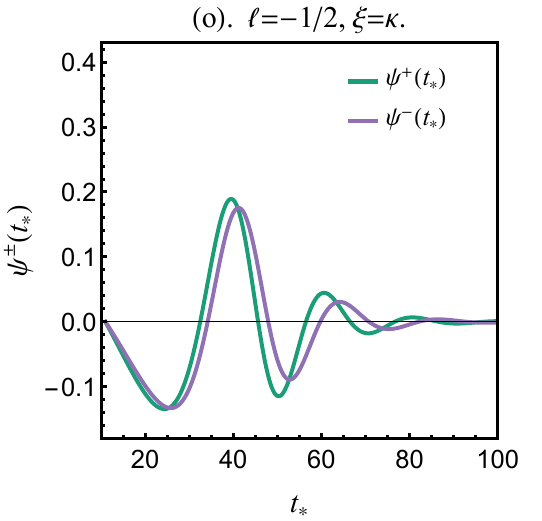}
\includegraphics[width=0.49\linewidth]{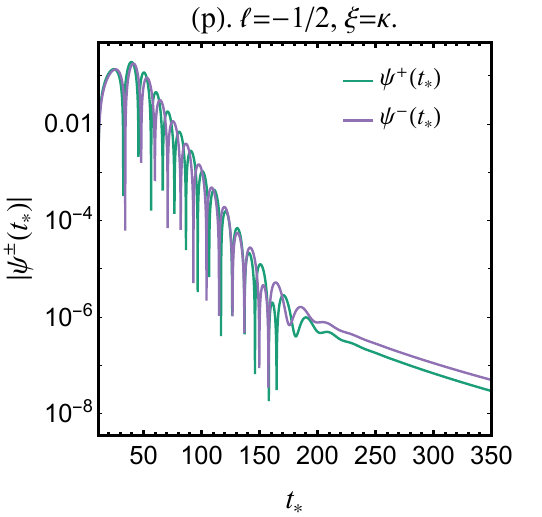}
\caption{The time evolution of the wave function $ \psi^{\pm}(t_*) $ corresponds to the vector perturbations in the $ l=2, n=0 $ mode.}
\label{fig4}
\end{figure}

In Figs. \ref{fig3}-\ref{fig4}, we present the ringdown waveforms corresponding to the azimuthal indices $l=1$ and $l=2$, generated with the black hole parameters $ M=1$ and $ \Lambda_e=10^{-4} $.
It can be noted that for non-zero Lorentz violation parameters, there is a significant inconsistency between the axial and polar sections waveforms.
With a positive Lorentz violation parameter, the polar modes exhibit a lower frequency than the axial modes.
However, their damping is nearly identical, implying that their QNMs have smaller real parts and identical imaginary parts. 
On the contrary, with negative parameters, corresponding to the "$-$" sign within the potential term $ V(B^\mu B_\mu \pm b^2) $ in the action \meq{Action}, the reverse situation occurs. 
It should be noted that the sign $ \pm $ indicates whether $ B_\mu $ is timelike ($ \ell=\xi b^2  $) or spacelike ($ \ell=-\xi b^2  $). 
These observations are in agreement with the conclusions drawn in the previous subsection. 
Furthermore, isospectral breaking is more obvious under negative parameters, and larger azimuthal indices $l$ tend to lessen the breaking, as they increasingly dominate the behavior of the perturbation effective potential.

An additional point of focus is the verification of the accuracy of the numerical method. 
This is done by attempting to fit $ \psi(t,r_*) $ using a QNM model comprising a finite number of exponentially damped sinusoids.
Without loss of generality, only considering the fundamental modes, we drop the indices $n$ and $m$.
We employ a modified exponentially decaying function,
\begin{align}
Q(t)=e^{\sqrt{1+\ell}\omega_I t}A_{l}\sin(\sqrt{1+\ell}\omega_R t+B_{l}), && t\in (t_0,t_\text{end}),
\end{align}
with a fitting range from $ t_0=120M $ to $ t_\text{end}=160M $.
Last, we present the fitted results in comparison with those calculated by the matrix method in Tabs. $ \ref{tab1} $ and $ \ref{tab2} $, which is the axial and polar sectors, respectively.
\begin{table}[h]
\renewcommand{\arraystretch}{1.2}
\centering
\caption{Contrasting the axial modes for values of $ M = 1 $, $ \Lambda_e = 0.01 $ and $ \xi=0.2\kappa $.}
\label{tab1}
\setlength\tabcolsep{2mm}{
\begin{tabular}{cccccc}
\hline\hline
\multirow{2}{*}{~\bf }	&
\multirow{1}{*}{~\bf $ \ell $~}  &
\multicolumn{1}{c}{Matrix method} & \multicolumn{1}{c}{Fitting}
\\
\hline
\multirow{3}{*}{$ l=1 $}
  &$0   $  &0.237424-0.088396i  &0.237668-0.0883007i           \\
  &$0.5 $  &0.243300-0.071398i	&0.243292-0.0714096i           \\
  &$1.0 $  &0.242528-0.060609i	&0.242522-0.0606198i           \\
\\
\multirow{3}{*}{$ l=2 $}
  &$0   $  &0.436871-0.090683i 	&0.436854-0.0907212i
\\&$0.5 $  &0.432678-0.072507i	&0.432683-0.0725695i
\\&$1.0 $  &0.424538-0.061270i	&0.424563-0.0613548i
\\
\hline\hline
\end{tabular}}
\end{table}
\begin{table}[h]
\renewcommand{\arraystretch}{1.2}
\centering
\caption{Contrasting the polar modes for values of $ M = 1 $, $ \Lambda_e = 0.01 $ and $ \xi=0.2\kappa $.}
\label{tab2}
\setlength\tabcolsep{2mm}{
\begin{tabular}{cccccc}
\hline\hline
\multirow{2}{*}{~\bf }	&
\multirow{1}{*}{~\bf $ \ell $~}  &
\multicolumn{1}{c}{Matrix method} & \multicolumn{1}{c}{Fitting}
\\
\hline
\multirow{3}{*}{$ l=1 $}
  &$0   $  &0.237424-0.088396i  &0.237668-0.0883007i           \\
  &$0.5 $  &0.238885-0.071341i	&0.238874-0.0713523i            \\
  &$1.0 $  &0.236195-0.060559i	&0.236188-0.0605673i             \\
\\
\multirow{3}{*}{$ l=2 $}
  &$0   $  &0.436871-0.090683i 	&0.436854-0.0907212i
\\&$0.5 $  &0.430204-0.072501i	&0.430208-0.0725615i
\\&$1.0 $  &0.420947-0.061264i	&0.420971-0.0613467i
\\
\hline\hline
\end{tabular}}
\end{table}
Despite the inherent errors that can arise during numerical computations and the limitations imposed by the finite number of parameters in the fitting process, we are confident in the accuracy of both the MM and the CFM, as evidenced by our obtained fitting results.

\section{CONCLUSIONS AND EXTENSIONS} \label{Sec.4}

In this paper, we investigate the QNMs of an Einstein-Bumblebee black hole solution with a cosmological constant. 
Here, the bumblebee field, characterized by a pure radial vacuum energy expectation, is given by $b_{\mu}=(0,b_r,0,0)$. 
Note that the bumblebee field can take forms such as $b_{\mu}=(0,b_r,b_\theta,0)$ \cite{Chen2020} or $b_{\mu}=(b_t,b_r,0,0)$ \cite{shao2022,Mai:2024lgk}.
For this solution, discussed in our paper, we derive the second-order Schr\"{o}dinger-like equation for axial and polar sections sourced by the vector field perturbation, and then solve its QNM spectrum using the Matrix Method and the Continued Fraction Method.
Our results show a significant dependence on the Lorentz violation parameter $ \ell $, and confirm that the nonzero vacuum expectation value of the Bumblebee field will lead to isospectrality breaking of the vector modes.
This breaking is attributed to the Lorentz violation parameter $ \ell $ in the background metric, which has an additional effect on the axial part of the term $  A^\mu R_{\mu\nu}\xi/\kappa $ in the perturbation equation.
In terms of $ \omega $, we observed, for the fundamental mode, the following trends:
\begin{itemize}[leftmargin=*,noitemsep,topsep=0pt,partopsep=0pt]
\item When both the Lorentz violation parameter $ \ell $ and the coupling constant $ \xi $ are non-zero, isospectrality breaking will occur. This effect becomes more pronounced as any of these parameters increases.
If considering a spacelike bumblebee field, compared to the timelike field, it introduces a more pronounced breaking effect.
\item As the azimuthal index $ l $ increases, it determines the dominant term of the frequency, leading to a reduced contribution from the $ \ell\xi/\kappa $ term in the axial effective potential. 
Therefore, it is observed that the breaking of isospectrality becomes increasingly less apparent.
\item Considering the Lorentz violation parameter $ \ell $, for a fixed azimuthal index $ l=1 $, the real part of QNMs first rises and then falls, while the magnitude of the imaginary part consistently decreases. However, for azimuthal indices $ l\geq2 $, both the real and imaginary parts increase and decrease monotonically, respectively.
\item The positive cosmological constant, consisting of the Lorentz violation parameter, has an effect similar to the Schwarzschild-dS, as referenced in \cite{Zhidenko:2003wq}.
\end{itemize}

Exploring the isospectrality issue of gravitational perturbations presents an interesting aspect within the Einstein-Bumblebee gravity theory.
In the first order approximation of the spacetime Lorentz violation parameter, it is possible to consider the trivial solution of the vector perturbations while maintaining the self-consistency of the gravitational perturbation equations.
The isospectrality breaking is foreseeable: axial perturbations naturally decouple from the bumblebee field, as described by the Regge-Wheeler equation \cite{ReggeWheeler1957,Liu:2022dcn}; for polar perturbations, the complexity of component field equations exceeds that of the Schwarzschild scenario, rendering decoupling via the conventional Zerilli method seemingly impossible \cite{Zerilli1970PRD,Zerilli1970PRL}.
The coupling of polar perturbations with the bumblebee field implies the breaking of isospectrality.
How to obtain the QNMs for polar modes in Einstein-Bumblebee gravity theory remains an open question.



\section{ACKNOWLEDGMENTS}
The authors gratefully acknowledge Songbai Chan, Qiyuan Pan and Qin Tan for insightful discussions.
This work was partially supported by the National Natural Science Foundation of China under Grants No. 12375046, No. 12122504 and No. 12035005.

\appendix


\begin{thebibliography}{000}
\bibitem{Kostelecky1998}
D. Colladay D and V. A. Kosteleck, \textit{Lorentz-violating extension of the standard model.} Phys. Rev. D, \textbf{58}, 116002 (1998).

\bibitem{Kostelecky1991}
V. A. Kosteleck and R. Potting, \textit{CPT and strings.} Nucl. Phys. B \textbf{359}, 545 (1991).

\bibitem{Kostelecky19891}
V. A. Kosteleck and S. Samuel, \textit{Phenomenological gravitational constraints on strings and higher-dimensional theories.} Phys. Rev. Lett. \textbf{63}, 224 (1989).

\bibitem{Gambini1999}
R. Gambini and J. Pullin, \textit{Nonstandard optics from quantum space-time.} Phys. Rev. D, \textbf{59}, 124021 (1999).

\bibitem{Kostelecky2001}
S. M. Carroll, J. A. Harvey and V. A. Kosteleck, et al. \textit{Noncommutative field theory and Lorentz violation.} Phys. Rev. Lett. \textbf{87}, 141601 (2001).

\bibitem{Kostelecky1989}
V. A. Kosteleck and S. Samuel, \textit{Gravitational phenomenology in higher-dimensional theories and strings.} Phy. Rev. D \textbf{40}, 1886 (1989).

\bibitem{Casana2018}
R. Casana, A. Cavalcante, F.P. Poulis and E.B. Santos, \textit{Exact Schwarzschild-like solution in a bumblebee gravity model}, Phys. Rev. D \textbf{97}, 104001 (2018).

\bibitem{Maluf2021}
R. V. Maluf and J. C. S. Neves, \textit{Black holes with a cosmological constant in bumblebee gravity.} Phys. Rev. D \textbf{103}, 044002 (2021).

\bibitem{Ovgun2019}
A. Ovgun, K. Jusufi and I. Sakall, \textit{Exact traversable wormhole solution in bumblebee gravity.} Phys. Rev. D \textbf{99}, 024042 (2019).

\bibitem{Gullu2020}
I. Gullu and A. Ovgun, (2020), \textit{Schwarzschild Like Solution with Global Monopole in Bumblebee Gravity}, arXiv:2012.02611 [gr-qc]


\bibitem{Ding2022}
C. Ding, X. Chen and X. Fu, \textit{ Einstein-Gauss-Bonnet gravity coupled to bumblebee field in four dimensional spacetime}. Nucl. Phys. B, \textbf{975}, 115688 (2022).

\bibitem{Ding2020}
C. Ding and C. Liu, et al. \textit{Exact Kerr-like solution and its shadow in a gravity model with spontaneous Lorentz symmetry breaking.} Eur. Phys. J. C \textbf{80}, 178 (2020).


\bibitem{Poulis:2021nqh}
F. P. Poulis and M. A. C. Soares, \textit{Exact modifications on a vacuum spacetime due to a gradient bumblebee field at its vacuum expectation value}, Eur. Phys. J. C \textbf{82}, 613 (2022).

\bibitem{Liu:2022dcn}
W.~Liu, X.~Fang, J.~Jing and J.~Wang, \textit{QNMs of slowly rotating Einstein-Bumblebee black hole}, Eur. Phys. J. C \textbf{83}, 83 (2023).

\bibitem{Mai:2023ggs}
Z. F. Mai, R. Xu, D. Liang and L. Shao, \textit{Extended thermodynamics of the bumblebee black holes}, Phys. Rev. D \textbf{108}, 024004 (2023). 

\bibitem{Yang:2023wtu}
K. Yang, Y. Z. Chen, Z. Q. Duan and J. Y. Zhao, \textit{Static and spherically symmetric black holes in gravity with a background Kalb-Ramond field}, [arXiv:2308.06613 [gr-qc]].

\bibitem{Xu:2023xqh}
R. Xu, D. Liang and L. Shao, \textit{Bumblebee Black Holes in Light of Event Horizon Telescope Observations}, Astrophys. J. \textbf{945}, 148 (2023).

\bibitem{Zhang:2023wwk}
X. Zhang, M. Wang and J. Jing, \textit{Quasinormal modes and late time tails of perturbation fields on a Schwarzschild-like black hole with a global monopole in the Einstein-bumblebee theory}, [arXiv:2307.10856 [gr-qc]].

\bibitem{Lin:2023foj}
R.~H.~Lin, R.~Jiang and X.~H.~Zhai, \textit{Quasinormal modes of the spherical bumblebee black holes with a global monopole}, Eur. Phys. J. C \textbf{83}, 720 (2023). 

\bibitem{Chen:2023cjd}
C.~Chen, Q.~Pan and J.~Jing, \textit{Quasinormal modes of a scalar perturbation around a rotating BTZ-like black hole in Einstein-bumblebee gravity}, [arXiv:2302.05861 [gr-qc]].

\bibitem{Chen2020}
S.B. Chen, M.Z. Wang, and J.L. Jing, \textit{Polarization effects in Kerr black hole shadow due to the coupling between photon and bumblebee field}, J. High Energy Phys. \textbf{1}, 17 (2020).

\bibitem{wang2021}
Z. Wang, S. Chen and J. Jing, \textit{Constraint on Lorentz symmetry breaking in Einstein-bumblebee theory by quasi-periodic oscillations.} preprint arXiv: 2112.02895, 2021.

\bibitem{Jing2022}
J. Jing, S. Chen, M. Sun, X. He, M. Wang and J. Wang, \textit{Self-consistent Effective-one-body theory for spinless binaries based on post-Minkowskian approximation I: Hamiltonian and decoupled equation for $\psi_B^4$} Sci. China-Phys. Mech. Astron. \textbf{65}, 260411 (2022).

\bibitem{Chandbook}
S. Chandrasekhar, \textit{The Mathematical Theory of Black Holes} (Oxford University Press, Oxford, 1992).

\bibitem{Pani2012}
P. Pani and V. Cardoso, et al. \textit{Perturbations of slowly rotating black holes: massive vector fields in the Kerr metric}, Phys. Rev. D \textbf{86}, 104017 (2012).

\bibitem{Pani2012prl}
P. Pani and V. Cardoso, et al. \textit{Black-hole bombs and photon-mass bounds}, Phys. rev. lett. \textbf{109}, 131102 (2012).

\bibitem{Pani2013prd}
P. Pani, E. Berti and L. Gualtieri, \textit{Scalar, electromagnetic, and gravitational perturbations of Kerr-Newman black holes in the slow-rotation limit}, Phys. Rev. D, \textbf{88}, 064048 (2013).

\bibitem{Pani2013prl}
P. Pani, E. Berti and L. Gualtieri, \textit{Gravitoelectromagnetic perturbations of Kerr-Newman black holes: stability and isospectrality in the slow-rotation limit}, Phys. Rev. Lett. \textbf{110}, 241103 (2013).

\bibitem{Pani2013IJMPA}
P. Pani, \textit{Advanced Methods in Black-hole Perturbation Theory} Int. J. Mod. Phys. A \textbf{28}, 22n23, 1340018 (2013).

\bibitem{Berti2009}
E. Berti and V. Cardoso, \textit{Quasinormal modes of black holes and black branes}. Class. Quant. Grav. \textbf{26}, 163001 (2009).

\bibitem{isoLoop}
D. del-Corral and J. Olmedo, \textit{Breaking of isospectrality of quasinormal modes in nonrotating loop quantum gravity black holes.} Phys. Rev. D \textbf{105}, 064053 (2022).

\bibitem{isoChSi}
S. Bhattacharyya and S. Shankaranarayanan, \textit{Distinguishing general relativity from Chern-Simons gravity using gravitational wave polarizations.} Phys. Rev. D \textbf{100}, 024022 (2019).

\bibitem{Isi:2019aib}
M. Isi, M. Giesler, W. M. Farr, M. A. Scheel and S. A. Teukolsky, \textit{Testing the no-hair theorem with GW150914}, Phys. Rev. Lett. \textbf{123}, 111102 (2019). 

\bibitem{Bluhm:2007bd}
R.~Bluhm, S.~H.~Fung and V.~A.~Kostelecky, \textit{Spontaneous Lorentz and Diffeomorphism Violation, Massive Modes, and Gravity}, Phys. Rev. D \textbf{77}, 065020 (2008). 

\bibitem{Thorne1980}
K. S. Thorne, \textit{Multipole Expansions of Gravitational Radiation} Rev. Mod. Phys. \textbf{52}, 299 (1980).

\bibitem{Thompson2016}
J. E. Thompson, B. F. Whiting and H. Chen, \textit{Gauge invariant perturbations of the Schwarzschild spacetime} Class. Quant. Grav. \textbf{34}, 174001 (2017).

\bibitem{guage2022}
W. T. Liu, X. J. Fang, J. L. Jing and A. Z. Wang, \textit{Gauge Invariant Perturbations of General Spherically Symmetric Spacetimes}.  Sci. China-Phys. Mech. Astron. \textbf{66}, 1-14 (2023).

\bibitem{Liu:2023uft}
W. Liu, X. Fang, J. Jing and J. Wang, \textit{Gravito-electromagnetic perturbations of MOG black holes with a cosmological constant: Quasinormal modes and Ringdown waveforms}, JCAP, \textbf{11}, 057 (2023).

\bibitem{Berti:2005eb}
E. Berti and K. D. Kokkotas, \textit{Quasinormal modes of Kerr-Newman black holes: Coupling of electromagnetic and gravitational perturbations}, Phys. Rev. D \textbf{71}, 124008 (2005).

\bibitem{Lin20171}
K. Lin and W.L. Qian, \textit{A matrix method for quasinormal modes: Schwarzschild black holes in asymptotically flat and (anti-) de Sitter spacetimes}. Class. Quant. Grav. \textbf{34}, 095004 (2017).

\bibitem{Lin2017}
K. Lin and W.L. Qian, et. al. \textit{A matrix method for quasinormal modes: Kerr and Kerr-Sen black holes}. Mod. Phys. Lett. A \textbf{32}, 1750134 (2017).

\bibitem{Lin2019}
K. Lin and W.L. Qian, \textit{The matrix method for black hole quasinormal modes}. Chinese Phys. C \textbf{43}, 035105 (2019).

\bibitem{Lei2021}
Y.H. Lei, M.J. Wang and J.L Jing, \textit{Maxwell perturbations in a cavity with Robin boundary conditions: two branches of modes with spectrum bifurcation on Schwarzschild black holes}. Eur. Phys. J. C \textbf{81}: 1-12 (2021).

\bibitem{Leaver1985}
E. W. Leaver, \textit{An analytic representation for the quasi-normal modes of Kerr black holes}.  Proc. Roy. Soc. Lond. \textbf{A402}, 285 (1985).

\bibitem{Yoshida:2003zz}
S. Yoshida and T. Futamase,
\textit{Numerical analysis of quasinormal modes in nearly extremal Schwarzschild-de Sitter space-times}. Phys. Rev. D \textbf{69}, 064025 (2004). 

\bibitem{shao2022}
R. Xu, D. Liang and L. Shao, \textit{Static spherical vacuum solutions in the bumblebee gravity model.} arXiv:2209.02209 (2022).

\bibitem{Mai:2024lgk}
Z. F. Mai, R. Xu, D. Liang and L. Shao, \textit{Dynamic instability analysis for bumblebee black holes: the odd parity}, [arXiv:2401.07757 [gr-qc]].

\bibitem{Zhidenko:2003wq}
A. Zhidenko, \textit{Quasinormal modes of Schwarzschild de Sitter black holes}, Class. Quant. Grav. \textbf{21}, 273-280 (2004).

\bibitem{ReggeWheeler1957}
T. Regge and J. A. Wheeler, \textit{Stability of a Schwarzschild singularity}, Phys. Rev. \textbf{108}, 1063 (1957).

\bibitem{Zerilli1970PRL}
F. J. Zerilli, \textit{Effective potential for even parity Regge-Wheeler gravitational perturbation equations}, Phys. Rev. Lett. \textbf{24}, 13: 737 (1970).

\bibitem{Zerilli1970PRD}
F. J. Zerilli, \textit{Gravitational field of a particle falling in a schwarzschild geometry analyzed in tensor harmonics}, Phys. Rev. D \textbf{2}, 10: 2141 (1970).
























\end{thebibliography}
\end{document}